\documentstyle[prl,aps,twocolumn,psfig,epsfig,floats]{revtex}
\begin{document}
\twocolumn[\hsize\textwidth\columnwidth\hsize\csname@twocolumnfalse\endcsname
\title{
Superconducting single-charge transistor in a tunable dissipative environment}
\author{Frank K. Wilhelm$^{1,2,3}$,
Gerd Sch\"on$^{1,4}$,
Gergely T. Zim\'anyi$^2$}

\address{$^1$Institut f\"ur Theoretische Festk\"orperphysik,
Universit\"at Karlsruhe, D-76128 Karlsruhe, Germany. \\
$^2$Department of Physics, UC Davis, Davis CA 95616, U.S.A\\
$^3$Quantum Transport Group, TU Delft,
P.O. Box 5046, 2600 GA Delft, The Netherlands\\
$^4$Forschungszentrum Karlsruhe, Institut f\"ur Nanotechnologie,
D-76021 Karlsruhe, Germany}

\maketitle
\begin{abstract}
  We study a superconducting single-charge transistor, where the
  coherence of Cooper pair tunneling is destroyed by the coupling to a 
  tunable dissipative environment. Sequential tunneling and  
  cotunneling processes are analyzed to construct the shape of the 
  conductance peaks and their dependence on the dissipation and
  temperature. Unexpected features are found due to a cross-over between
  two distinct regimes, one `environment-assisted' the other
  `environment-dominated'. Several of the predictions have been
  confirmed by recent experiments. The model and results apply also to
  the dynamics of Josephson junction quantum bits on a conducting ground
  plane, thus explaining the influence of dissipation on the coherence. 
  
\end{abstract}
\pacs{PACS numbers: 74.50.+r, 74.25.Fy}
]

The behavior of a quantum system coupled to a dissipative 
environment is one of the paradigms of modern physics 
\cite{Leggett}. It is the central challenge in any attempt to
build and manipulate quantum information systems, since fluctuations
and dissipation limit the quantum coherence of the device. For
instance, for qubits based on two different charge states of a 
Josephson junction~\cite{Makhlin} one important source of dissipation
are the normal electrons in the ground plane. A conducting ground
plane is 
an appropriate tool
however,  to compensate for the random offset charges
of the superconducting island. In the experimental realization   
of Nakamura et al.~\cite{Nakamura} this was achieved by mounting the device
on a gold ground plane.

\begin{figure}[hbt]
\centering
(a) \epsfig{file=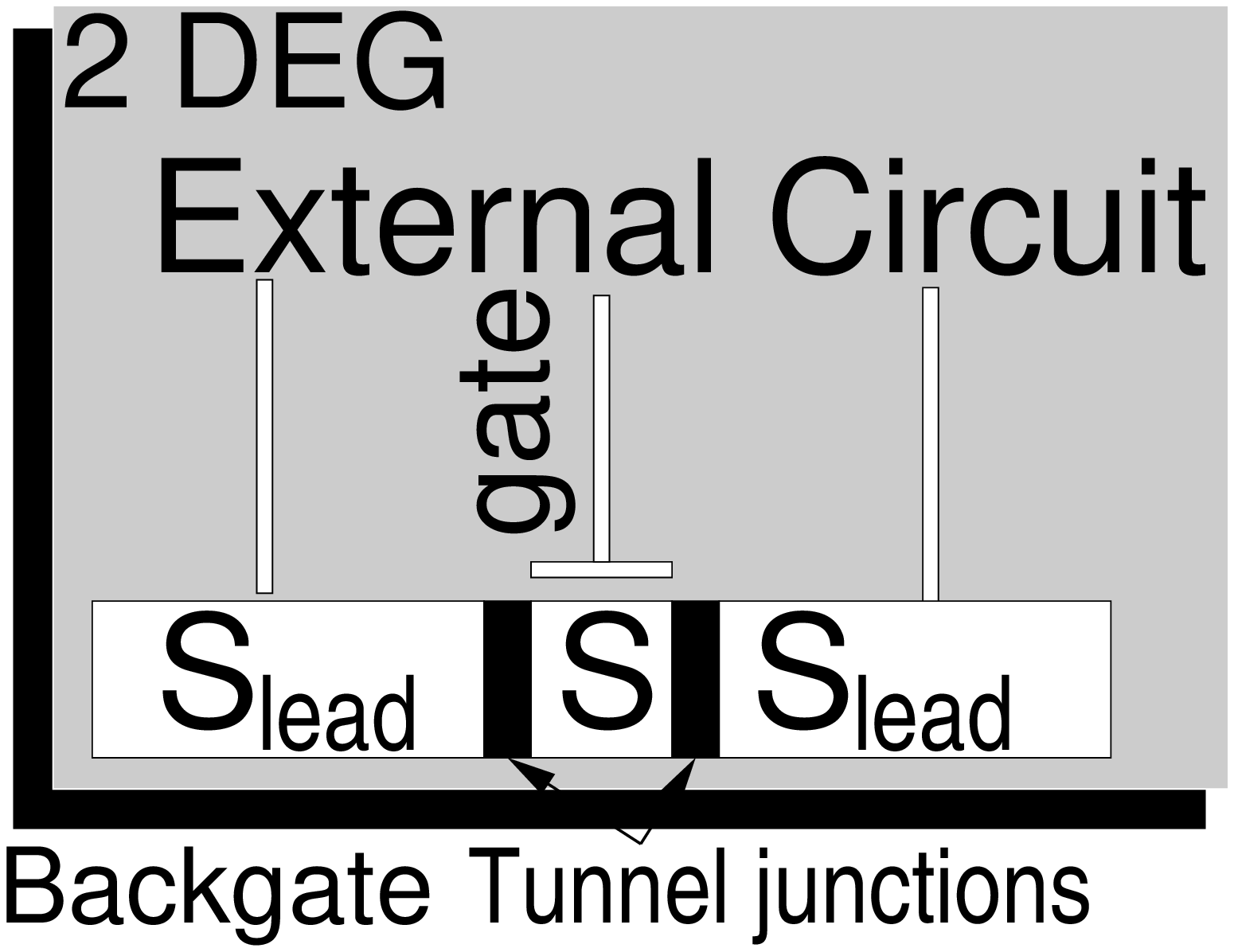,width=0.42\columnwidth}
(b) \epsfig{file=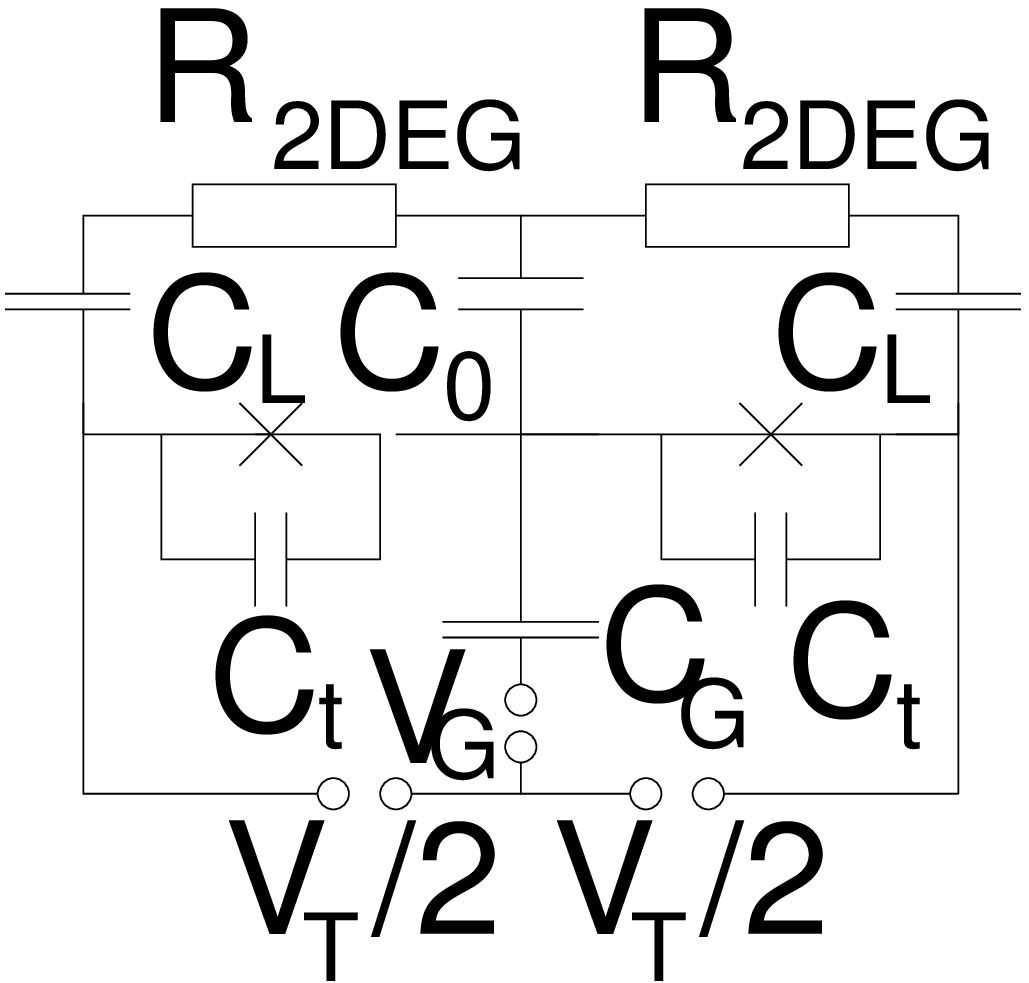,width=0.42\columnwidth}
\vspace{2mm}
\caption{(a) Top view of the system under consideration. The 2DEG is
in the substrate.  (b) Equivalent circuit.}
\label{fig1}
\end{figure}

In earlier experiments the Berkeley group~\cite{berkeley1} had 
mounted an array of Josephson junctions on a 
two-dimensional electron gas (2DEG), separated from it by an
insulating layer. In this setup the  density of normal electrons in the 2DEG,
and hence the source and strength of the dissipation, can be tuned by
a back gate over a wide range (see also~\cite{Kuzmin}).  
This influences the collective properties of the array in an qualitative
way~\cite{wagenblast}. In more recent experiments~\cite{berkeley}
the Berkeley group investigated the transport properties of a
Superconducting Single-Charge Transistor (SSCT) coupled to a tunable
2DEG (see Fig.~\ref{fig1}). They observed an unexpected dependence of
the conductance peak heights and widths on temperature and dissipation.

In previous work the nonlinear 
current
branch of a SSCT
has been studied~\cite{matveev,joyez,paul}. 
Due to the  coupling of the SSCT to the metallic ground plane 
 the Cooper pair charges interact with normal metal image charges,
which turns the tunneling of Cooper pairs dissipative.  
A central result of the present paper is 
the identification and characterization of the previously uninvestigated 
normal current branch in the {\it linear} and nonlinear  response regime. 
We determine how the conductance depends on the 
gate and transport voltage, environment conductance and temperature.
We also provide estimates how  dissipation
limits the quantum oscillations of qubits.

In the following we concentrate on systems (see Fig.~\ref{fig1}) with
low resistance per square of the ground plane,   
$R\ll  h/4e^2 \approx 6.5\,$k$\Omega$, and we choose the 
gate and junction capacitances, $C_{\rm g}$ and $C_{\rm j}$, 
and the capacitances between the island and the leads to the 
metallic plane, $C_0$ and $C_{\rm l}$, to be ordered as follows:
$C_{\rm g}$, $C_{\rm j} \ll C_0 \ll C_{\rm l}$. 
The effects of the dissipative metallic ground plane on the SSCT are
captured by the real part of the impedance, 
\begin{equation}
\mbox{Re}\{Z(\omega)\}
= R/[1+(\omega RC_{\rm j})^2] \, .
\label{ReZ}
\end{equation} 
Interestingly, even though the different capacitances introduce several 
frequency regimes, in all regimes Re$Z(\omega)$ is determined by the
capacitance of the tunnel junction $C_{\rm j}$ only.   
Quasiparticle dissipation may be ignored at temperatures 
much below the superconducting gap. 

We construct our theory following Ref.~\cite{PEDevoret,Gra-Dev}. 
The environment can be described by a harmonic oscillator bath. An
integration over the quadratic bath degrees of freedom is possible.
The resulting sequential tunneling rate through one
junction~\cite{ANO},
\begin{equation}
        \Gamma(\delta E_{\rm ch}) = (\pi/2\hbar) E_{\rm J}^2 
        {P}(\delta E_{\rm ch}) \; ,
\label{Cooperrate}
\end{equation}
depends on the change in charging energy during the
tunneling process, which in turn
depends on the gate voltage $n_{\rm g} = V_{\rm g}C_{\rm g}/e$ 
and the transport voltage $V_{\rm tr}$. For a process which
increases the number of excess Cooper pairs from $n$ to $n+1$ is given by
\begin{equation}
\delta E_{\rm ch}=-4E_{\rm C}(2n-n_{\rm G}+1) + eV_{\rm tr} ~~.
\end{equation}
The (total) capacitance of the island determines the energy scale
$E_{\rm C}=e^2/2(2C_{\rm j} + C_{\rm g})$, while
$E_{\rm J}$ is the Josephson coupling energy of the tunnel junctions
(here assumed to be equal).   
The function $P(E)$ can be expressed by a correlation function
$ K(t)=$ 
$\frac{4e^2}{h}\int_{-\infty}^{\infty} \frac{d\omega}{\omega} 
        \mbox{Re}\{Z(\omega)\}
        \left\{ \coth{\big(\frac{\hbar \omega}{2 T} \big) }
 [\cos(\omega t) - 1] 
        - i \sin(\omega t) \right\}
$
via
$
        {P}(E) = \frac{1}{2\pi \hbar} \int_{-\infty}^{\infty}
        d t     \exp\left[{K}(t) + i \frac{Et}{\hbar}\right]
$.

In the situation considered, $K(t)$ can be evaluated analytically 
and in the long-time limit, $|t|\gg RC_{\rm j}$, it  reads 
\begin{eqnarray}
K(t)&=& -\frac{2}{{g}}\left[\pi T|t|
+\log\left(1-e^{-2\pi T|t|}\right)\right.\nonumber\\
&&+\left.\gamma+\log(2\pi TRC_{\rm j})+i\hbox{sign}(t) \frac{\pi}{2}\right].
\label{Ktapprox}
\end{eqnarray}
In the low energy regime, $E,T\ll RC_{\rm j}$, we find
\begin{equation}
{P}(E)=\frac{(2\pi e^{-\gamma}TRC_{\rm j})^{\frac{2}{g}}}{2\pi^2T}
\hbox{Re}\left[e^{-i\frac{\pi}{g}}
B\left(\frac{1}{g}-\frac{iE}{2\pi T},1-\frac{2}{g}\right)
\right]\label{PofE}
\end{equation}
where $\gamma=0.577 \dots$, ${g}=h/4e^2R$ is the dimensionless
conductance of the ground plane,  and $B(x,y)$  the Beta function.
We observe that the energy appears only in the dimensionless 
combination $\epsilon=Eg/2\pi T$.
\begin{figure}
\epsfig{file=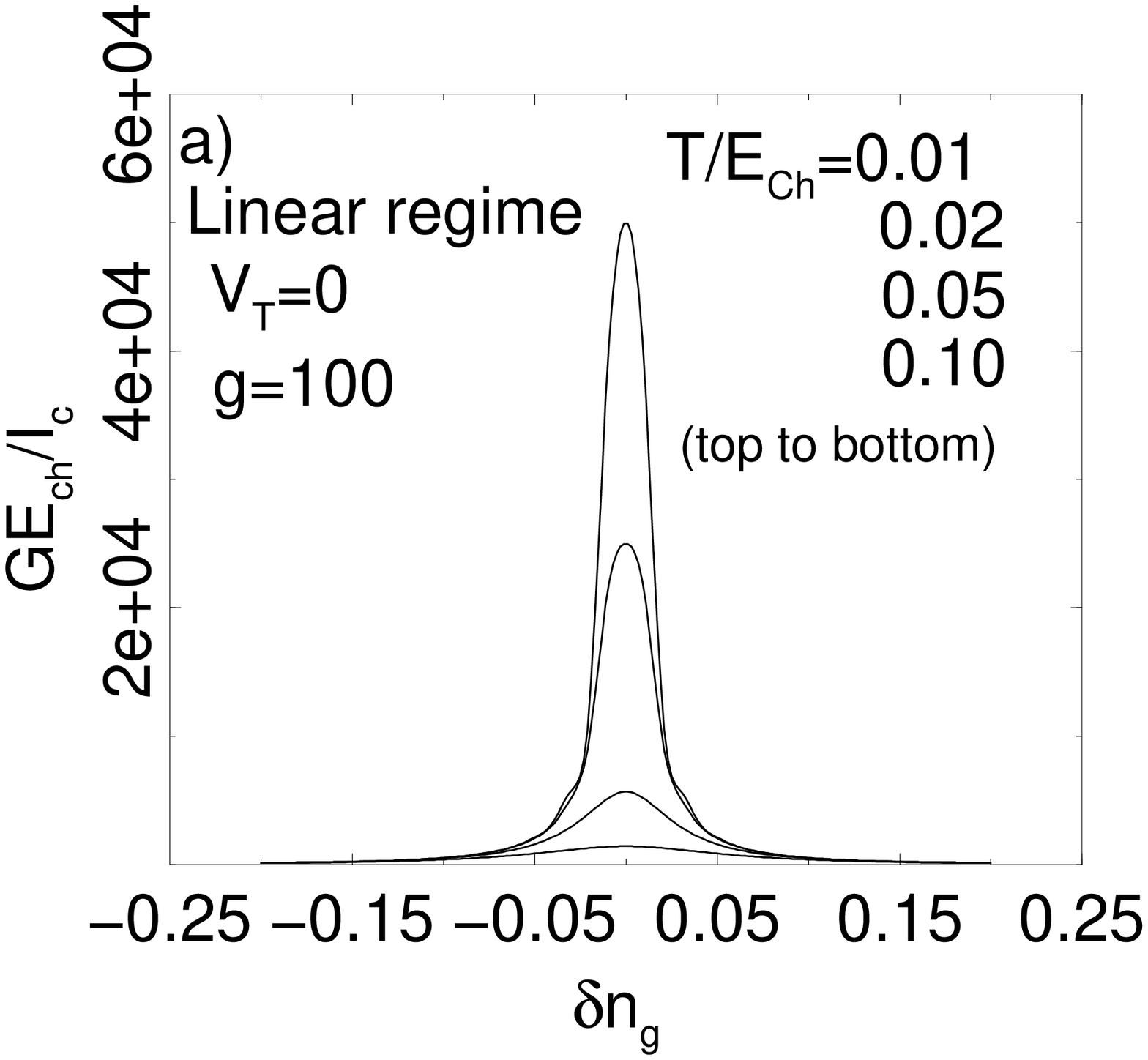,width=0.48\columnwidth}
\hfill
\epsfig{file=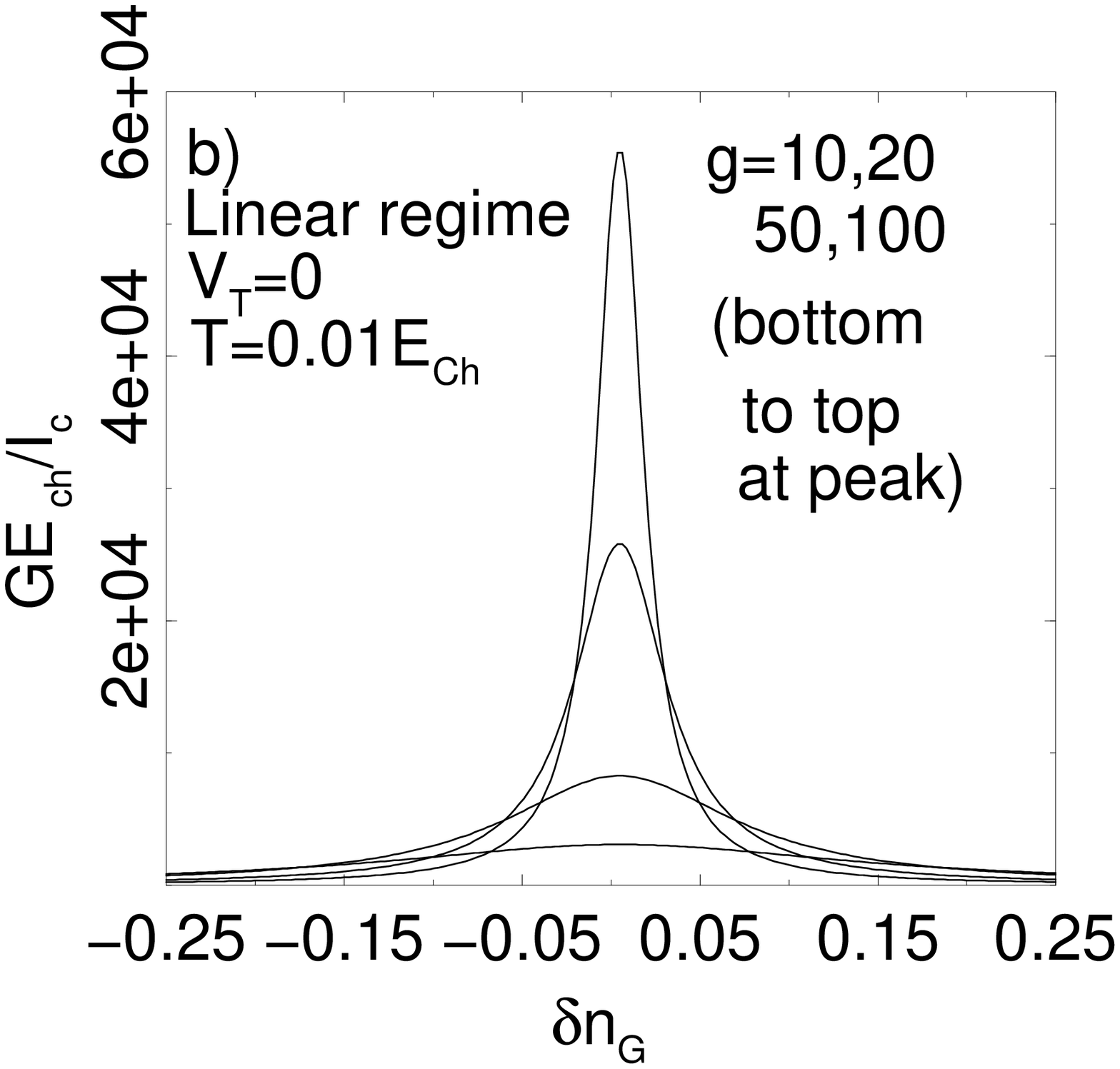,width=0.48\columnwidth}
\epsfig{file=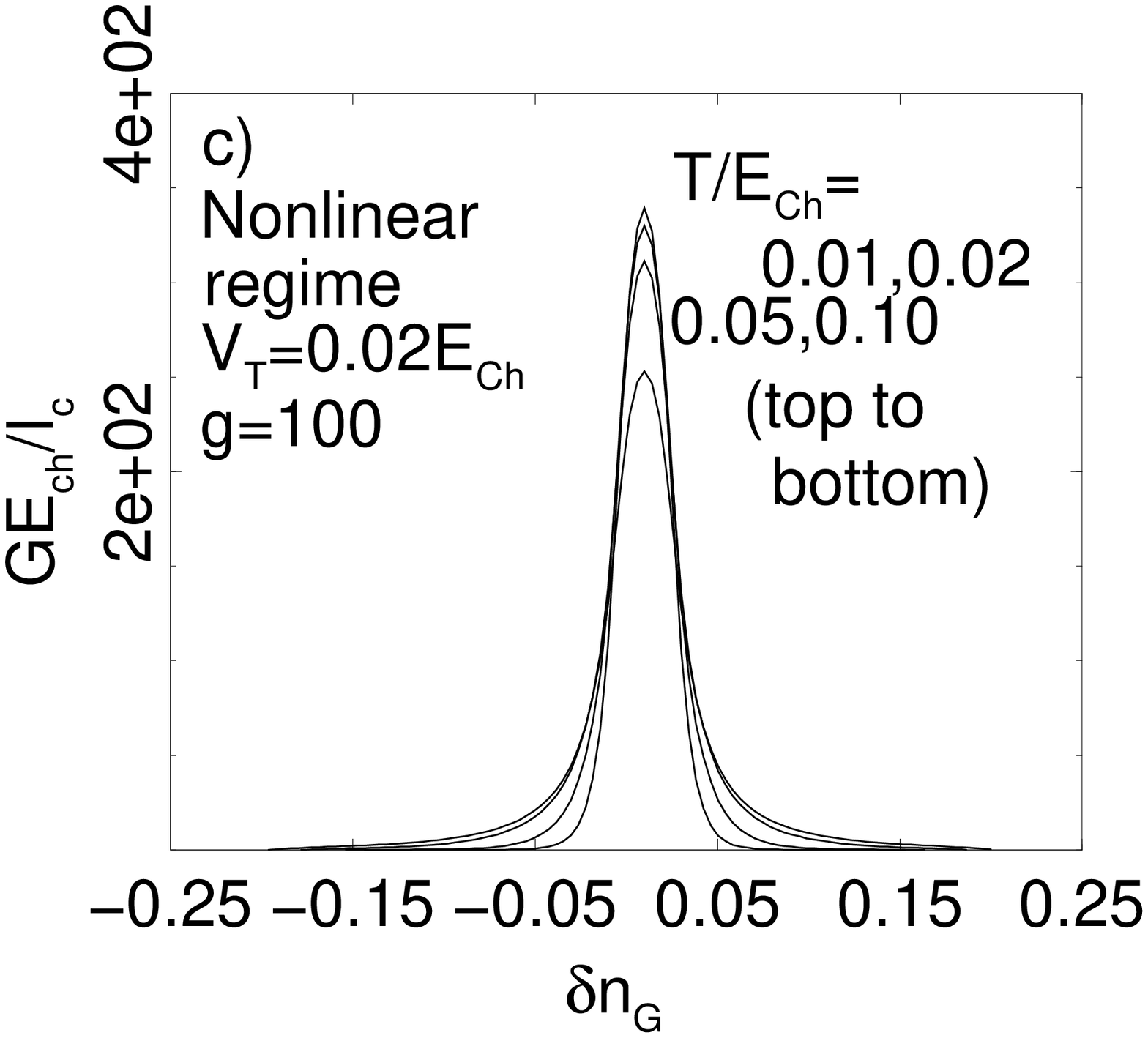,width=0.48\columnwidth}
\hfill
\epsfig{file=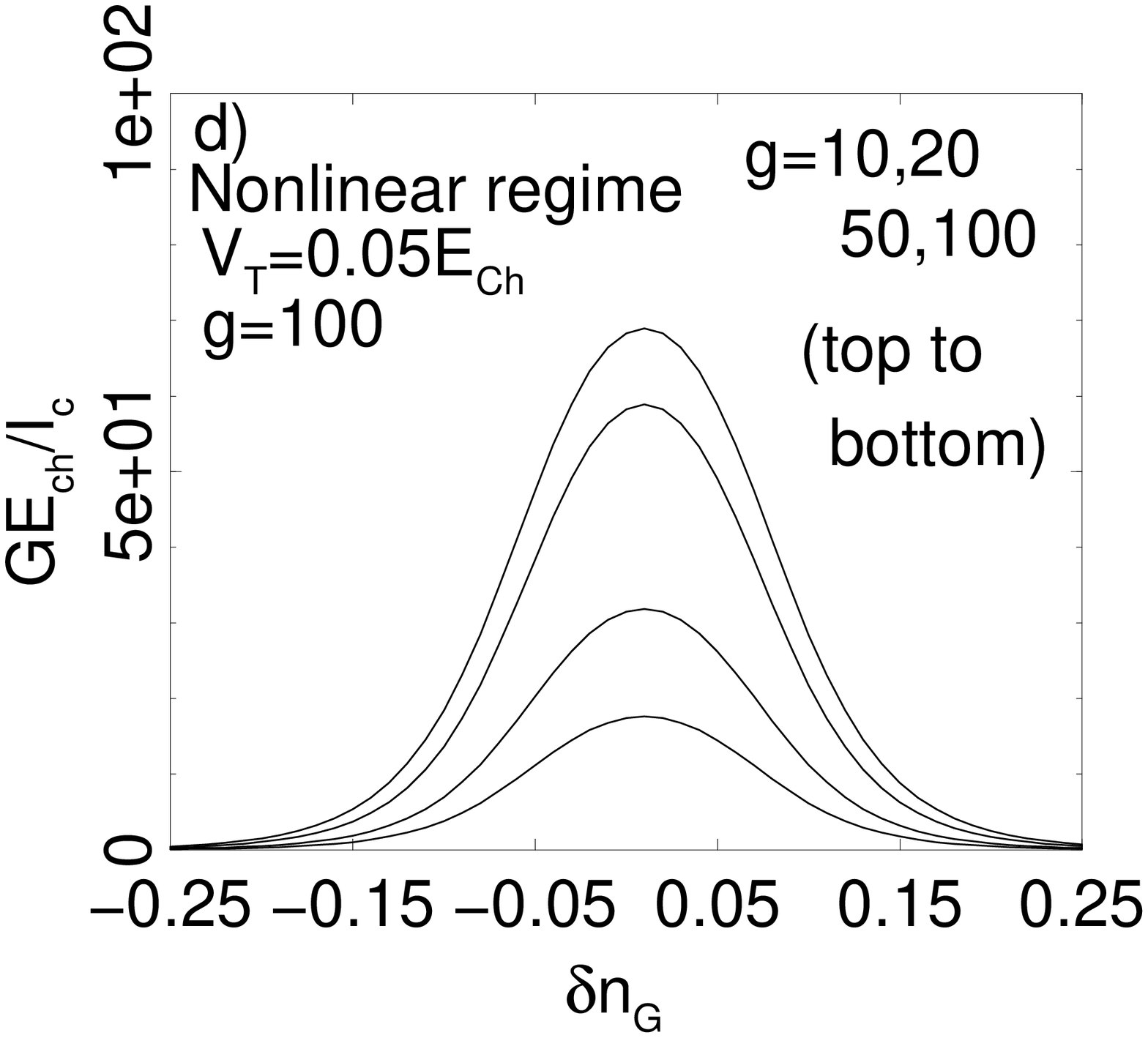,width=0.48\columnwidth}
\caption{Conductance peaks in lowest order (sequential tunneling)
scaled by the critical current of a single junction, $I_c$ and 
the charging energy $E_{\rm C}$ of the island, assuming 
$C_{\rm G}\ll C_{\rm j}$.}
\label{peaks}
\end{figure}

Given the tunneling rates 
$\Gamma(\delta E_{\rm ch})$  we construct a master 
equation and determine the current $I$ through the SSCT.
The difference of the charging energies $\delta E_{\rm ch}$ varies as the 
gate voltage is tuned. At low temperatures, for most values of $n_{\rm
g}$, the large cost in charging energy suppresses transport by the Coulomb
blockade. At the degeneracy points of the states with $n$ and $n+1$
Cooper pairs (i.e. $\delta E_{\rm ch}=0$), transport
becomes possible. Thus the conductance
exhibits Coulomb oscillations as a function of the gate voltage.

Previously published work~\cite{Gra-Dev,Schoen} concentrated on the
low temperature regime  $\epsilon \gg 1$. Here the asymptotic expansion of
Eq.~(\ref{PofE}) yields ${P}(E)\approx 2/{g} E$
\cite{PEDevoret}. For instance at the degeneracy point, we have
$\delta E_{\rm ch}=eV_{\rm tr}$ and the $I$-$V$ characteristics is
strongly nonlinear 
\begin{equation}
I = 4\pi E_{\rm J}^2/\hbar gV_{\rm tr} \, .
\label{I0}
\end{equation}
Away from the degeneracy point the high energy cost,  
$\delta E_{\rm ch} \gg T$, suppresses 
tunneling (Coulomb blockade). Thus this regime is still described by 
$|\epsilon| \gg 1$. Detailed balance requires
$P(-E) = e^{-E/ T} P(E)$, from which one finds that
the conductance is suppressed by an exponential activation
factor~\cite{PEDevoret}. 

A closer inspection shows, however, that for typical transport 
voltages of $V_{\rm tr}\approx 1\;\hbox{nV}$, even at the 
lowest temperatures of $T=20mK$, for the typical values of $g=1..100$
one has  $\epsilon \ll 1$.  
This motivated us to study this previously ignored parameter regime.
Surprisingly we find that both at the peaks and in the Coulomb blockade 
regime the system's response becomes linear, although for different 
reasons. 

First, {\it at the peaks} and for low transport voltages $\delta E_{\rm ch} =
eV_{\rm tr} \ll  T$ one can expand Eq.~($\ref{PofE}$). We find 
that in this regime the current response is linear 
\begin{equation}
I(V_{\rm tr},n_{\rm G}=n+1) = e\pi E_{\rm J}^2 geV_{\rm tr}
/\hbar T^2 \, , 
\label{IT}
\end{equation}
and the height of the conductance peak now {\em increases} with 
increasing conductance $g$ of the metallic plane,
in contrast to what could be concluded from Eq. (\ref{I0}).

The full shape of the conductance peak derived from Eq.~(\ref{PofE}) 
is shown in Figure 2. At low enough transport voltages, $V_{\rm tr} < 2\pi T/g$, 
{\it at the peaks} we have $\epsilon < 1$ and the current response is linear and
transport is provided by `environment-{\em dominated}' sequential
tunneling (ED-ST). The peak height is proportional to $1/T^2$ at fixed
$g$ (Fig.~\ref{peaks}(a)) and proportional 
to $g$ at fixed $T$ (Fig.~\ref{peaks}(b)). 
{\it Away from the peak} the energy change $\delta E_{\rm ch}$
for a tunneling process is large even for $V_{\rm tr}=0$.
Therefore one crosses over to the regime $\epsilon \gg 1$. Here the differential 
conductance at some fixed voltage is proportional to $1/g$. This tendency 
is opposite to that of the peaks, and manifests itself in the crossing of the 
conductance 
$G(V_{\rm g})$ curves in the {\it tails} (see Fig.~\ref{peaks}(b)). 
In this `environment-{\em assisted}' sequential tunneling (EA-ST)
regime, the conductance may still be appreciable; it is reduced due to
Coulomb blockade only for $\delta E_{\rm ch}>T$, i.e. at $\epsilon>g$. 
We also find that the width of the peak increases with decreasing $g$, and 
the conductance in the tails decreases with increasing $T$. 

In a transport experiment the voltage is 
necessarily finite, $V_{\rm tr} > 0$. At low
enough temperatures $T \ll V_{\rm tr} g / 2\pi $ this forces the
system to $\epsilon > 1$, 
where the nonlinear response  formulae govern the physics. 
Thus, the apparent divergence of the conductance {\it at the peak},
$G\simeq 1/T^2$ is regularized by the transition to the nonlinear
regime,  yielding a {\it finite} conductance for $T\rightarrow 0$ 
(Fig.~\ref{peaks}(c)).
Also, at 
sufficiently large
$V_{\rm tr}$ the conductance
monotonically  
increases with decreasing $g$ over the whole energy range 
(Fig.~\ref{peaks}(d)).

In second order in $E_{\rm J}/E_{\rm ch}$ 
the current is exponentially suppressed in the
Coulomb blockade regime. Here, higher order terms, such as the fourth order 
cotunneling process, may yield important contributions. Cotunneling
describes the simultaneous coherent transfer of two Cooper pairs,
where the the energetically forbidden intermediate state is occupied
only virtually. As a consequence the process is not suppressed by an
exponential  activation factor in the blockade regime. The
most relevant term (for $\delta E_{\rm ch}\ll -T/g$) 
leads to the rate~\cite{Odintsov}
$$
\Gamma_{\rm CT}=\frac{E_J^4}{2\pi \delta E_{\rm ch}^2
T}\hbox{Re}\left[e^{-4i\pi/g} 
B\left(-i\frac{V_{\rm tr}}{2\pi T}+\frac{4}{g},1-\frac{8}{g}\right)\right].
$$
At low voltages we thus find again a {\it linear response} regime
with a decay rate $\Gamma_{\rm CT}\propto V_{\rm tr}g/T^2$. 

Combining these results, we can read off the peak width within 
linear response. Depending on the experimental conditions, the 
conductance is dominated by environment-assisted sequential tunneling
(EA-ST) or environment-dominated cotunneling (ED-CT). This leads to
a half-width of  
$\delta E_{\rm ch,1/2}\propto\hbox{max}\left(T/g,E_{\rm J}\right)$ whereas
a fixed off-peak conductance $g_{\rm 0}$ is reached at $\delta
E_{\rm ch,g_0}\propto\hbox{max}\left(E_{\rm J}/\sqrt{gg_0},\sqrt{g/g_0}~E_{\rm
J}^2/T\right)$. 
At high transport voltages and large $\delta E_{\rm ch}$ the response 
becomes {\it nonlinear}, yielding $\Gamma_{\rm CT}\propto 1/g V^2$.
Remarkably however the ED-ST regime extents to 
large $V_{\rm tr}$, in the region where the bare charging energy difference 
and $V_{\rm tr}$ combine to a small net $\delta E_{\rm ch}$, generating
again a low value of $|\epsilon|\ll 1$. Here again the incoherent
transport of Cooper pairs dominates, with a conductance $G \propto g
E_J^2/T^2$. The expressions for the conductances in 
the different parameter regimes are summarized in Fig. 3.
\begin{figure}[htb]
\begin{center}
\epsfig{file=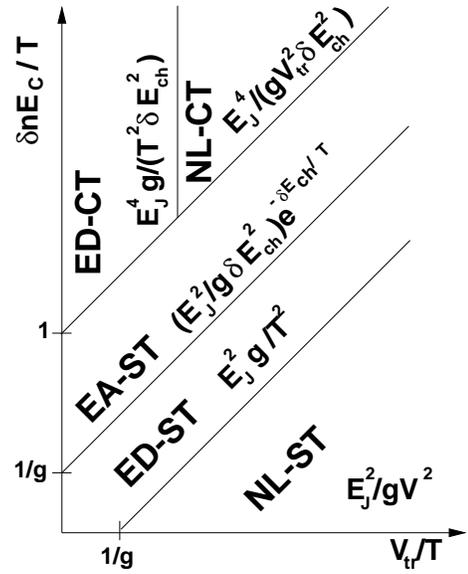,width=0.7\columnwidth}
\end{center}
\caption{Overview of the different transport
regimes as a function of the transport voltage
$V_{\rm tr}$ and the gate voltage (entering in the combination
$\delta E_{\rm Ch,0}=-(n-n_{\rm G}-1/2)4e^2/C$).
ED-CT: Environment-dominated cotunneling; NL-CT: Nonlinear cotunneling;
EA-ST: Environment-assisted sequential tunneling; ED-ST: Environment-dominated 
sequential tunneling; NL-ST: Nonlinear sequential tunneling}
\end{figure}

So far we considered relatively short and wide leads, when the bilayer 
formed by the leads and the metallic plane can be described as a capacitor. 
If the leads are longer and narrower, the spatial correlations 
of charges moving within the bilayer modify the fluctuation spectruma and
should be taken into
account. 
Then the proper model for the leads is an RC transmission line
using the per-square capacitance and resistance, 
rather than the $C_{\rm L}$-capacitors in Fig.~1.
The impedance of such lines is $Z_{RC}(\omega)=\sqrt{R/i\omega C_{\rm sq}}$
where $C_{\rm sq}$ is the square capacitance between the lead and the ground
plane. 
In this case the $\omega^{-1/2}$ divergence governs the
impedance.
At $T=0$ the model has been analysed before~\cite{Schoen,Gra-Dev}.
Here we concentrate on the high temperature limit.
{\it At the peak} and for $T\gtrsim E$ we expand in powers of 
$\omega/2T$ to obtain
$$
K(t)=-2\sqrt{\pi |t|/~gC_{\rm sq}} ~(4|t|T-i\hbox{sign}(t))
$$
which is dominated by the $|t|^{3/2}$ term. Then 
$$
P(0)= \Gamma(2/3)/3 \cdot \left(C_{\rm sq}g/2\pi T^2 e^2
\right)^{1/3}~~.
$$
The response of the system is again linear, but the condutance scales 
with new exponents: $G \propto T^{-5/3}$ and $G \propto g^{1/3}$.

If the SSCT is driven into the normal state, its own dynamics becomes
dissipative, but it is still strongly influenced by 
the environment. In contrats to the resonant tunneling situation of
the superconducting state we have to sum now over the {\it normal} final  
states. In this case the tunneling rate  is found from $P(E)$, eq. \ref{PofE}
along the lines of \cite{Gra-Dev}
\begin{eqnarray*}
\Gamma(\delta E_{\rm ch})&&= -\frac{2T}{R}
(e^{-\gamma} 2\pi TRC_{\rm j})^{2/g}\times\\
&&\times\hbox{Re}\left[e^{-i\pi/g}B\left(
1+\frac1g-i\frac{\delta E_{\rm ch}}{2\pi T},-1-\frac2g\right)\right] \;.
\end{eqnarray*}
It depends on the energy again through $\epsilon$. 
Since the charge carriers are now normal electrons, $g$ is redefined 
accordingly as $g=h/e^2R$ and the junction strength
is characterized by the tunneling resistance $R_{\rm t}$. 
For $\epsilon\ll 1$ we find  
$\Gamma_{\rm L}=(T/R_{\rm t})(1+\delta E_{\rm ch}/T)$ for
$g \rightarrow \infty$. Thus {\it at the peak} the response is linear
again, but this time with a temperature {\it independent} value as 
$T \rightarrow 0$.
For $\epsilon \gg 1$, the rate displays the usual Coulomb blockade
$$\Gamma (\delta E_{\rm ch})=\theta(\delta E_{\rm
  ch})\frac{\delta E_{\rm ch}g}{\pi
  R_{\rm t}\Gamma(2+2/g)}(e^{-\gamma}\delta E_{\rm ch}~RC_{\rm t})^{2/g} ~.$$
\begin{figure}
\epsfig{file=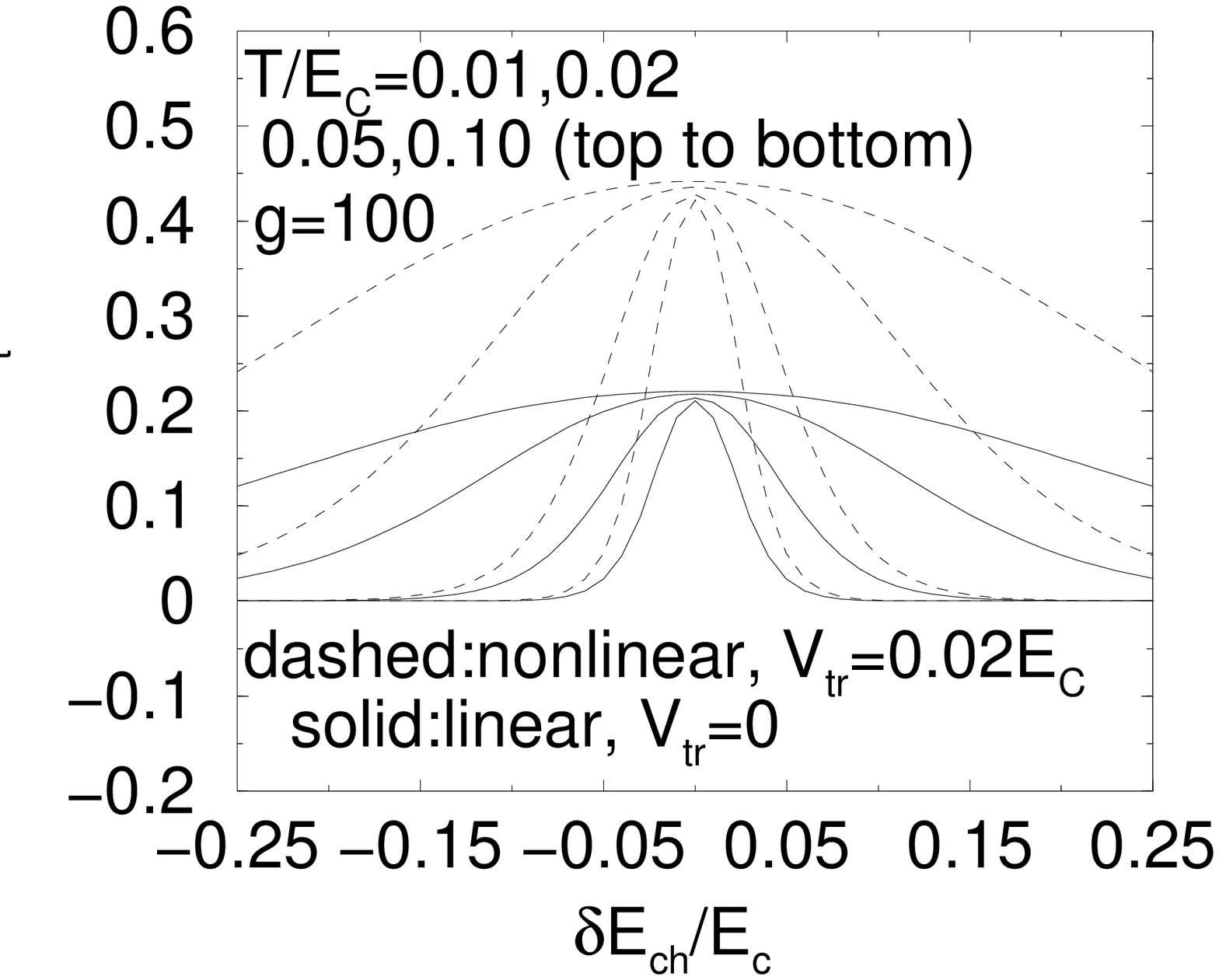,width=0.48\columnwidth}
\hfill
\epsfig{file=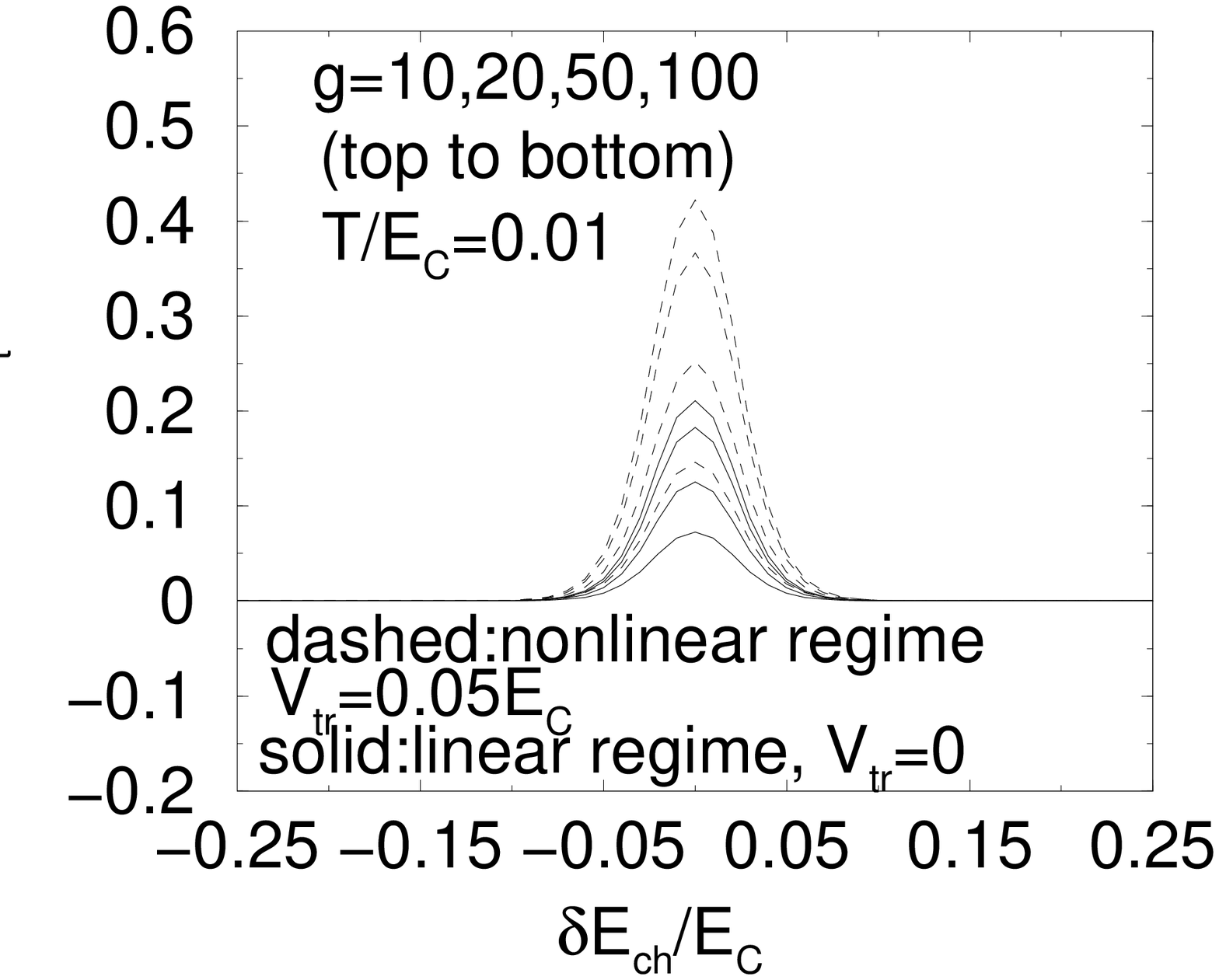,width=0.48\columnwidth}
\caption{Conductance peaks in lowest order of perturbation theory for the normal
state, scaled by the tunneling conductance $G_{\rm t}$ of the junctions.  \label{npeaks}}
\end{figure}
The change of the power-laws can be understood by the additional phase
space factor $\propto E^2/T^2$ in the normal state. The resulting
independence of the environment is caused by the fact that the system
is dissipative by itself and the external dissipation is weak at large
$g$. Our numerical results in Fig. \ref{npeaks} show the (weak) dependence
on the environment conductance in the $T=0$-limit. 

We are now ready to review the experimental results of the Berkeley
group~\cite{berkeley}. 
They constructed the SSCT just like that of Fig. 1. The
parameters were: $E_J = 1$ K, $C_t = 0.3 $fF,$ C_0 = 1.5 $fF.
The backplane conductance was swept in the range $g = 1 ... 100$
and the temperature in the range $T = 20 ... 1000$ mK. 
They found in the Coulomb oscillations of the conductance a linear 
response regime {\it at the peaks}, confirming a major prediction of the 
present theory. The peak conductance
was found to scale as $G \propto g^{0.3}$ and $G \propto T^{-0.9}$.
Their lead arrangement is modeled more adequately with a transmission line.
Our relevant results for this case are: $G \propto g^{1/3}$ and $G \propto 
T^{-5/3}$. When the SSCT was driven normal by a large magnetic field,
they found that the peak conductance exhibits very weak dependence on $g$
and $T$. Our theory predicted that $G$ goes to a constant for either
large $g$ or small $T$. So several of the  measured $g$ and $T$ 
dependences are in agreement with the present theory.
Finally, between the peaks they do {\it not} find an exponential
suppression of $G$, making it plausible that indeed the cotunneling
process is the dominant transport channel, as described above.

A similar system, a superconducting electron box mounted on a metallic
ground plane~\cite{NakamuraJLTP}, has been used to demonstrate the coherent
oscillations of charge states~\cite{Nakamura}, which makes this setup a 
promising candidate for a qubit in quantum computers. 
To this end, it has to be operated {\it at the peak}.
The quality factor $Q$ of the oscillations can be determined by standard
methods~\cite{Leggett}. At the low temperatures of $T\ll E_{\rm J}$ we
obtain
\begin{equation}
Q=\cot (\pi/(g-2))\approx g/\pi ~~,
\end{equation}
as for the metallic case, $g\gg 2$. For the parameters of~\cite{Nakamura} 
we find $Q$ in the range of several hundreds.
Thus the dissipation allows of the order of
a thousand cycles of the quantum oscillations. 
While this is encouraging, it still shows that
 the metal planes, which were introduced to
screen the random charges, induce a potentially serious limitation on
the maximum number of coherent operations.

Numerous stimulating discussions with J.\ Clarke, T. Giamarchi, L. Glazman, 
C.\ Kurdak, R.\ Therrien, J.B.\ Kycia, J.\ Chen, J.\ Siewert, G.\ Falci, 
J.\ K\"onig, and A.D. Zaikin are acknowledged. The work at UC Davis was was 
supported by NSF-9720440, NSF-DMR 9985978 and by NATO 971615. 
FKW and GS are supported by the DFG through SFB 195 and GK 284 and by the 
EU through TMR Superconducting Nano-circuits. 
GTZ's work at the UCSB ITP has been supported by  NSF-PHY-94-07194.

\end{document}